\documentclass[structabstract]{aa}

\usepackage{graphicx,xcolor,textpos}
\usepackage{rotating}
\usepackage{txfonts}
\usepackage{hyperref}
\usepackage[utf8]{inputenc}
\DeclareUnicodeCharacter{00A0}{~}
\bibpunct{(}{)}{;}{a}{}{,}

\begin{document}

\title{HD~41641: A classical $\delta$~Sct-type pulsator with\\
chemical signatures of an Ap star
\thanks{The CoRoT  space mission was developed and operated by the French space agency CNES, with participation of ESA's RSSD and Science Programmes, Austria, Belgium, Brazil, Germany, and Spain.}\fnmsep
\thanks{This work uses ground-based spectroscopic observations made with the HARPS instrument at the 3.6 m-ESO telescope (La Silla, Chile) under the Large Program 185.D-0056}
}

\author{A. Escorza\inst{1,2} \and
K. Zwintz\inst{1,3} \and
A. Tkachenko\inst{1} \and
T. Van Reeth\inst{1} \and
T. Ryabchikova\inst{4} \and
C. Neiner\inst{5} \and
E. Poretti\inst{6} \and
M. Rainer\inst{6} \and \\
E. Michel\inst{5} \and
A. Baglin\inst{5} \and 
C. Aerts\inst{1,7}
}

\offprints{A. Escorza, \\ \email{ana.escorza@ster.kuleuven.be}}

\institute{
    Institute of Astronomy, KU Leuven, Celestijnenlaan 200D, B-3001 Leuven, Belgium \\
    \email ana.escorza@ster.kuleuven.be \and
    Institut d'Astronomie et d'Astrophysique, Universit\'{e} Libre de Bruxelles, Boulevard du Triomphe, B-1050 Bruxelles, Belgium \and
    Institute for Astro-and Particle Physics, University of Innsbruck, Technikerstrasse 25/8, A-6020 Innsbruck, Austria \and
    Institute of Astronomy, Russian Academy of Sciences, Pyatnistkaya 48, 119017 Moscow, Russia \and
        LESIA, Observatoire de Paris, PSL Research University, CNRS, Sorbonne Universit\'{e}s, UPMC Univ. Paris 06, Univ. Paris Diderot, Sorbonne Paris Cit\'{e}, 5 place Jules Janssen, 92195 Meudon, France \and
    INAF-Osservatorio Astronomico di Brera, via E. Bianchi 46, I-23807 Merate (LC), Italy \and
    Department of Astrophysics/IMAPP, Radboud University Nijmegen, PO Box 9010, 6500 GL Nijmegen, The Netherlands
    }

\date{Received / Accepted }

\abstract
{Among the known groups of pulsating stars, $\delta$~Sct stars are one of the least understood. Theoretical models do not predict the oscillation frequencies that observations reveal. Complete asteroseismic studies are necessary to improve these models and better understand the internal structure of these targets.}
{We study the $\delta$~Sct star HD~41641 with the ultimate goal of understanding its oscillation pattern.}
{The target was simultaneously observed by the CoRoT space telescope and the HARPS high-resolution spectrograph. The photometric data set was analyzed with the software package PERIOD04, while FAMIAS was used to analyze the line profile variations. The method of spectrum synthesis was used for spectroscopically determining the fundamental atmospheric parameters and individual chemical abundances.}
{A total of 90 different frequencies was identified and analyzed. An unambiguous identification of the azimuthal order of the surface geometry could only be provided for the dominant p-mode, which was found to be a nonradial prograde mode with \textit{m}~=~+1. Using $T_\mathrm{eff}$ and \ensuremath{\log g}, we estimated the mass, radius, and evolutionary stage of HD~41641. We find HD~41641 to be a moderately rotating, slightly evolved $\delta$~Sct star with subsolar overall atmospheric metal content and unexpected chemical peculiarities.}
{HD~41641 is a pure $\delta$~Sct pulsator with p-mode frequencies in the range from 10~d$^{-1}$ to 20~d$^{-1}$. This pulsating star presents chemical signatures of an Ap star and rotational modulation due to surface inhomogeneities, which we consider indirect evidence of the presence of a magnetic field.}

\keywords{asteroseismology - stars: variables: $\delta$ Sct - stars: oscillations - stars: individual: HD 41641 - techniques: photometric - techniques: spectroscopic}

\titlerunning{HD~41641: A classical $\delta$~Sct-type pulsator with chemical signatures of an Ap star}
\authorrunning{Escorza et al.}

\maketitle

\section{Introduction}\label{sec:intro}

Although the stellar interior is not directly observable, asteroseismology allows us to unravel some of its details through the study of stellar oscillations. Detecting nonradial oscillations was not trivial from ground-based observations, but space missions such as MOST (Microvariability and Oscillations of STars; \citealt{MOST}), CoRoT (COnvection, ROtation et Transits plan\'{e}taires; \citealt{CoRoT}), and Kepler \citep{kepler}, revolutionized this branch of astronomy. Space photometry has been particularly revealing in the case of the asteroseismology of red giants (e.g., \citealt{RG_CoRoT}; \citealt{RG1}; \citealt{RG2}) and the derivation of internal rotation profiles of various types of stars (e.g., \citealt{ROT1}; \citealt{ROT2}).

\begin{figure*}[tb]
\centering
\includegraphics[width=0.75\textwidth,clip]{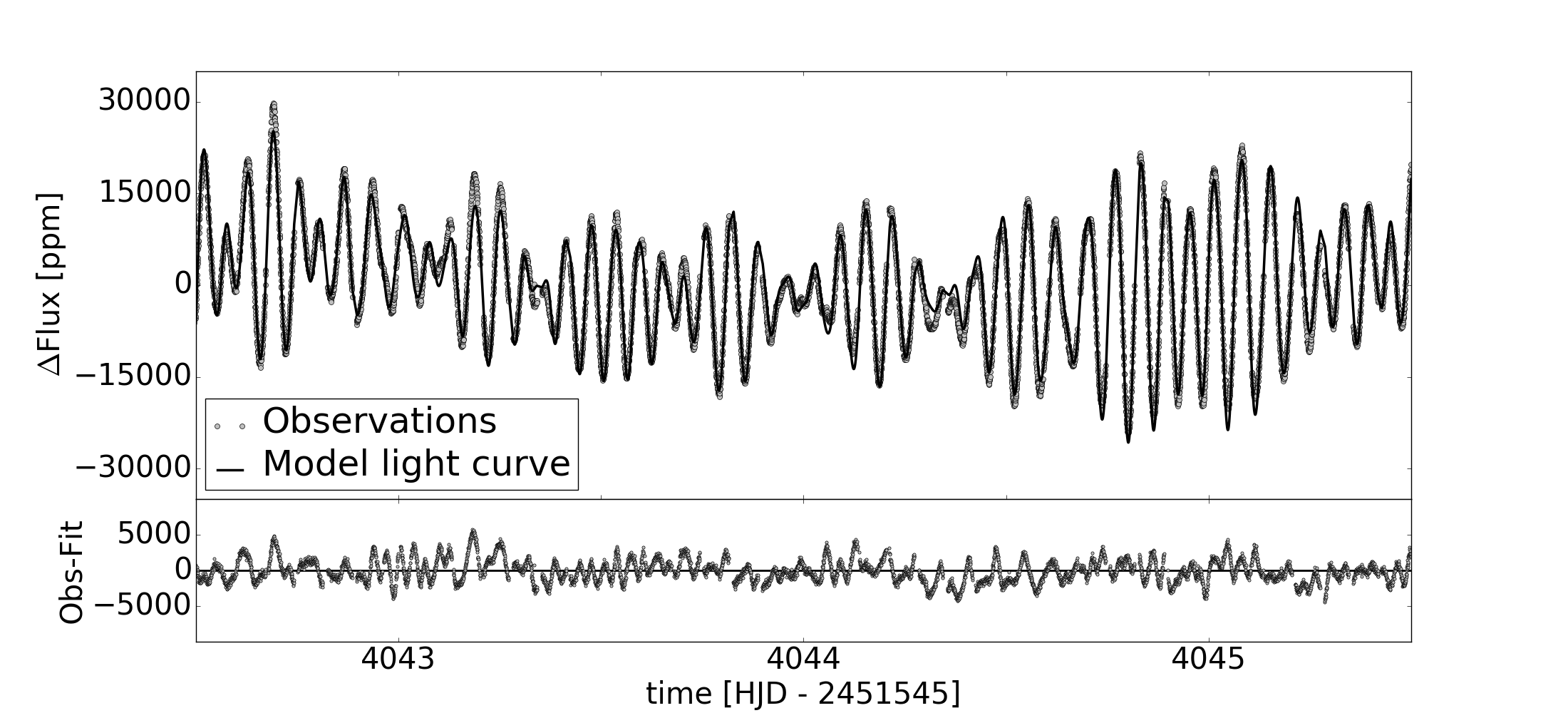}
\caption{Three-day subset of the CoRoT light curve of HD~41641 and model obtained with the 90 extracted frequencies.}
\label{LCzoom} 
\end{figure*}

The $\delta$~Sct instability region is located in the lower part of the classical instability strip, where it intersects with the main sequence. They are stars of spectral type A or F with masses in a range between 1.5 \textit{M}$_{\sun}$ and 4.0 \textit{M}$_{\sun}$ \citep{BOOK}. The instability strip of $\gamma$~Dor stars overlaps in the Hertzsprung-Russell diagram (HRD) with part of the $\delta$~Sct instability strip. Hybrid pulsators showing both $\delta$~Sct and $\gamma$~Dor pulsating properties have been found and are very interesting targets for asteroseismic studies \citep{hybrids}. In the region of the HRD where A- and F-type stars lie, different pulsation mechanisms and a wide variety of stellar processes (magnetic fields, convection, rotation...) can cause variability.

Among A-type stars, there is a group of chemically peculiar (CP) stars that show inhomogeneous surface distribution of some chemical elements. These inhomogeneities are referred to as chemical spots (e.g., \citealt{Lueftinger03}; \citealt{Lehmann07}). Most of these CP A-stars, called Ap stars, have global magnetic fields. Chemical spots in Ap stars may cause rotationally modulated light curves (e.g., \citealt{Krticka09}; \citealt{Shulyak10}). The coolest members of the Ap stars (the so-called roAp stars) have effective temperatures in a range of about 6600 to 8500~K \citep{roAp_T} and are pulsating with typical periods of 6 to 24 minutes (e.g., \citealt{Smalley15}). The Ap stars with the longest periods are generally the most evolved. All of these pulsating stars have overabundances of the rare-earth elements and show the so-called rare-earth anomaly, when the abundances derived from the lines of the second ions exceed those derived from the lines of the first ions by 1~dex or more \citep{Ryabchikova04}. The Nd-anomaly in the longest period roAp stars does not exceed 1~dex, however \citep{Alentiev12}.

In the HRD, roAp stars appear in the same instability strip as $\delta$~Sct stars. The pulsation periods of $\delta$~Sct stars range from about 20 minutes to 0.3 days. They are usually multiperiodic and present both radial and nonradial p-modes, generally of low degree \textit{l}. The amplitudes of those oscillation modes cover a wide range of several to several thousands parts-per-million (ppm) in flux. The $\kappa$~mechanism acting in the second ionization zone of helium is responsible for their pulsational behavior \citep{delta_pulsations}. While the pulsation periods of roAp and $\delta$~Sct stars partially overlap, only one $\delta$~Sct pulsator has been detected with a magnetic field. \cite{Kurtz08} claimed the first observational evidence for $\delta$~Sct type pulsations and magnetic field in an Ap star (HD~21190). However, \cite{Bagnulo12} considered both the magnetic field detection and the Ap classification of this star as spurious and further detailed spectroscopic analysis is necessary. The classical $\delta$~Sct star HD~188774 \citep{magdelta} is the only target  as yet clearly proven to exhibit magnetic signatures along with its pure  $\delta$~Sct pulsation properties.

Although $\delta$~Sct variables form one of the first groups of stars that was discovered to pulsate, it is still one of the least understood. Several hundreds of frequencies have been detected for some targets studied from space data, revealing much more frequencies than models predict (e.g., \citealt{Poretti09}; \citealt{deltaskepler}). Additionally, some of the theoretically expected frequencies are not detected observationally. There are some very well-studied $\delta$~Sct stars, such as FG Vir (e.g., \citealt{Breger09}) and 4 CVn (e.g., \citealt{Schmid14}), which have been targets of extensive photometric and spectroscopic observation campaigns. Despite the huge efforts, no seismic model has been achieved for these objects yet. Different $\delta$~Sct stars challenge the theory in different aspects, hence it is important to increase the number of well-studied targets.

HD 41641 ($\alpha_{2000}=6^{\mathrm{h}}6^{\mathrm{m}}40.578^{\mathrm{s}}$, $\delta_{2000}=$ +6\degr43'49.886'') was simultaneously observed by the CoRoT space telescope \citep{CoRoT} and the HARPS high-resolution spectrograph \citep{HARPS}. This bright star ($V=7.86$ mag, \citealt{intro3}), which has a spectral type of A5III \citep{intro2} and a radial velocity of 28.60 $\pm$ 2.50 $\mathrm{km\,s}^{-1}$ \citep{intro4}, has never been analyzed before from an asteroseismic point of view. The variability of HD~41641 was discovered in the preparatory work of the CoRoT mission \citep{ThisStar}. The fundamental parameters were obtained from $uvby\beta$ photometry, and HD~41641 resulted to be a $\delta$ Sct star located well inside the instability strip. A possible SB2 nature was also suspected from a high-resolution ELODIE spectrum.

In this study, we combine space photometry of HD~41641 with ground-based spectroscopy (Sect. \ref{sec:obs}) to conduct an asteroseismic investigation. The atmospheric parameters and individual chemical abundances of the target are determined for the first time from spectroscopic data, allowing us to discuss HD 41641's evolutionary stage and some interesting chemical peculiarities (Sect. \ref{sec:param}). A frequency analysis is performed on both data sets and spectroscopic mode identification is also discussed (Sect. \ref{sec:freqana}).

\section{Observations and data reduction}\label{sec:obs}

\subsection{Photometric data from CoRoT}\label{ssec:CoRoT}

The CoRoT space telescope was launched on December 27, 2006, aboard a Soyuz rocket into a low Earth polar orbit at an altitude of 896~km \citep{CoRoT}. The satellite carried a 27 cm diameter telescope equipped with a four CCD camera, which provided continuous and long observations of the same stellar fields. These fields of size 1.4\degr$\times$2.8\degr are situated within two cones of 10\degr of diameter, centered at right ascensions \textit{RA} = 18h50m and \textit{RA} = 6h50m, which are sometimes referred to as the \textit{CoRoT eyes}. CoRoT is no longer operational since a hardware failure in November 2012.

HD~41641 was observed continuously by CoRoT for 94 days, from December 17, 2010 to March 22, 2011, during the Long Run LRa05 as one of the five targets of the bright targets field (ex seismo field). After having removed the invalid measurements, most of which had been acquired when the satellite was crossing the South Atlantic Anomaly (SAA), the light curve consists of 223\,146 good measurements with a duty cycle of 87.5$\%$. Table~\ref{infophot} summarizes the characteristics of the data set, and Fig.~\ref{LCzoom} shows a three-day subset of the CoRoT light curve. The model light curve, obtained with all the extracted frequencies (see Sect.~\ref{ssec:CoRoTfreq}), is overplotted, and the residuals are shown in the lower panel.

\begin{table}[tb] 
\begin{small}
\caption{Characteristics of the CoRoT observations of HD~41641.}
\label{infophot}
\begin{center}
\begin{tabular}{cccccc}
\hline\hline
\rule[0mm]{0mm}{3mm}
CoRoT & Valid  & Duty & length, T & 1/T & f$_{Nyquist}$\\
\rule[0mm]{0mm}{2mm}
run & points & cycle & [d] & [d$^{-1}$] & [d$^{-1}$]\\
\hline
\rule[0mm]{0mm}{4mm}
LRa05 & 223\,146 & 87.5$\%$ & 94 & $\sim$ 0.0106 & 1320.91\\
\hline\hline
\end{tabular}
\end{center}
\end{small}
\tablefoot{Name of the CoRoT run, number of valid flux points used for the analysis, duty cycle, temporal length of the data set (T), Rayleigh frequency resolution (1/T), and Nyquist frequency (f$_{Nyquist}$).}
\end{table}

\subsection{Spectroscopic data from HARPS}\label{ssec:HARPS}

To complement the photometric observations by CoRoT, HD~41641 was observed in the high-efficiency mode (EGGS, $R$=80\,000) of the \'{e}chelle spectrograph HARPS (High Accuracy Radial velocity Planet Searcher; \citealt{HARPSnew}) at the 3.6 m telescope at ESO La Silla (Chile). In total, 222 spectra were taken during 15 nights, between December 23, 2010 and January 12, 2011, overlapping with part of the CoRoT observations. They cover the whole spectral range of HARPS, from 3780~\text{\AA} to 6914~\text{\AA}, and their exposure times range from 1000 seconds to 1200 seconds. The spectra were reduced using a semi-automated pipeline developed at the Brera observatory \citep{Rainer03}. During this process, they were flat-field corrected, deblazed using the continuum of a hot star, wavelength calibrated, and the barycentric correction was applied.

Additionally, a careful normalization was conducted following the procedure developed in \cite{normPapics}. With this method, a cubic spline function is fitted to the continuum, the latter defined through a careful selection of a few tens of points in the wavelength regions free of spectral lines. This was performed with a semiautomatic script, which also takes care of the order merging. It takes the signal-to-noise (S/N) values in the overlapping ranges into account and corrects for the sometimes slightly different flux levels of the overlapping orders.

\section{Fundamental parameters and chemical analysis}\label{sec:param}

The average normalized spectrum was analyzed with the  Grid Search in Stellar Parameters (GSSP) code \citep{GSSPnew} with the purpose of determining HD~41641's fundamental parameters, i.e., effective temperature, $T_\mathrm{eff}$; surface gravity, \ensuremath{\log g}; metallicity, $[M/H]$; projected rotational velocity, \ensuremath{{\upsilon}\sin i}; and microturbulent velocity, $\xi$. The code compares the observed spectrum with a series of synthetic spectra computed in a grid of the five considered parameters. It looks for the combination of these parameters that minimizes the difference between the synthetic and observed spectra in a $\chi^2$ minimization procedure. The code makes use of a grid of atmospheric models (see \citealt{GSSP} for details on the grid), which had been precomputed with the LLmodels software package \citep{LLmodels}. The synthetic spectra are computed on the fly by means of the SynthV radiative transfer code \citep{SynthV}. Both the atmosphere model and the radiative transfer code can handle individual elemental abundances and are based on the assumption of local thermodynamical equilibrium (LTE), which is appropriate for the temperature regime dealt with here.

\begin{figure}[tb]
\centering
\includegraphics[width=0.49\textwidth,clip]{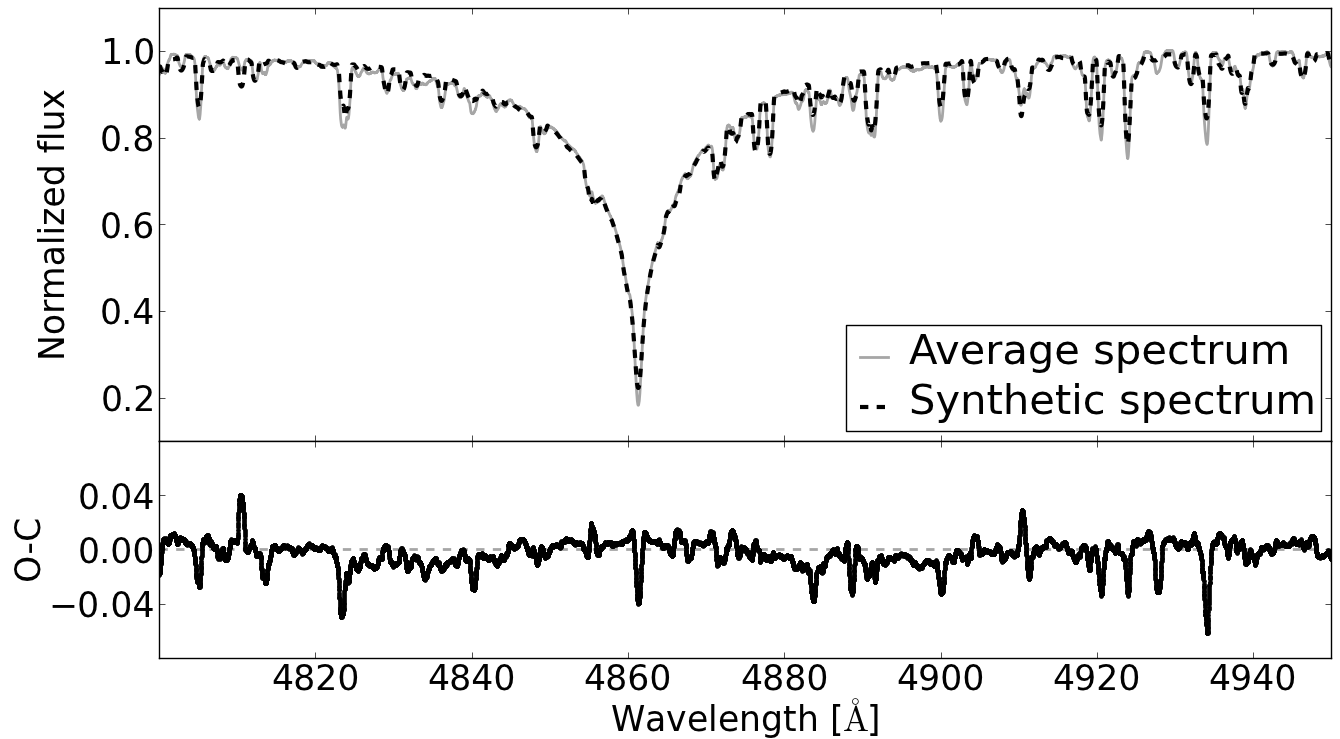}\\
\includegraphics[width=0.49\textwidth,clip]{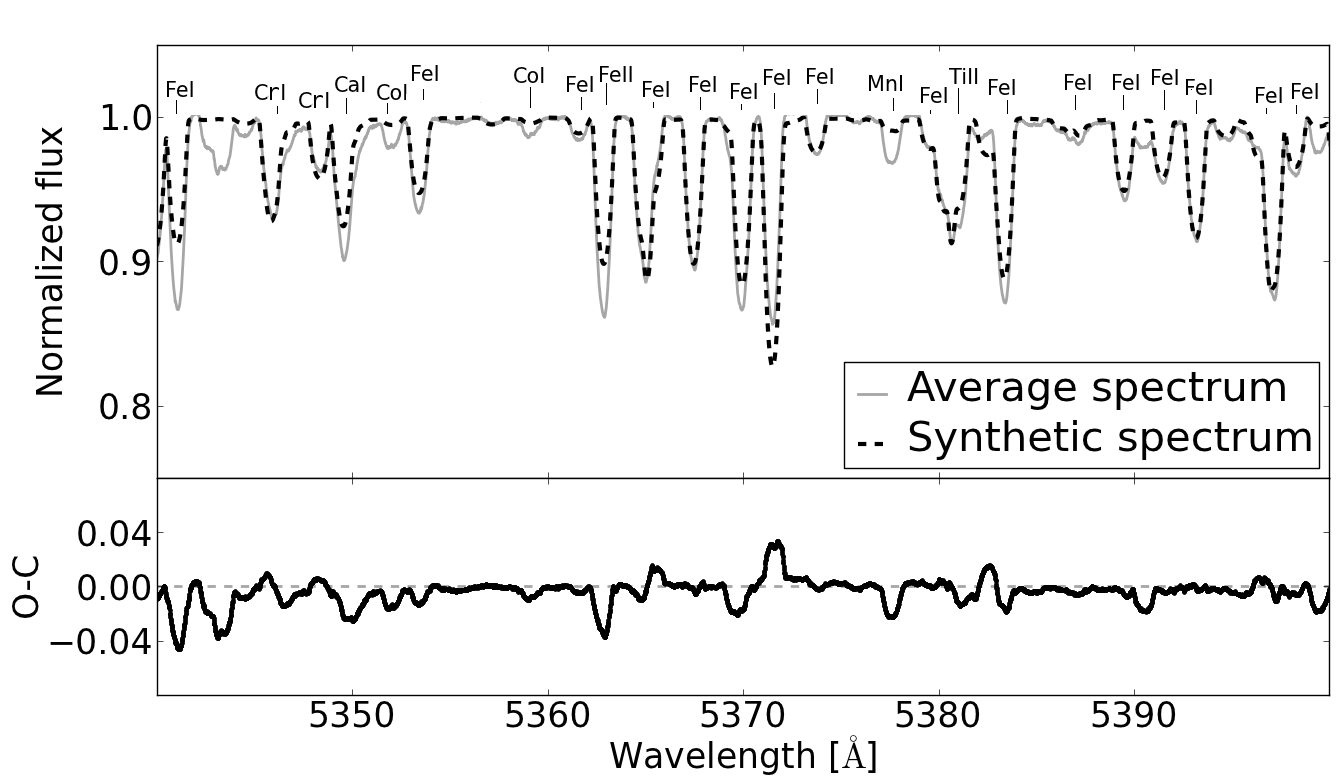}
\caption{\label{GSSPfit} Fit of the synthetic spectrum (dashed black line) of the H$\beta$ line (upper plot) and a group of metal lines (lower plot) to the average spectrum (solid gray line).}
\end{figure}

The wavelength range analyzed was from 4700~\text{\AA} to 5700~\text{\AA}, which includes the $H\beta$ line ($\lambda$ = 4861.323 \text{\AA}) and a wide region of metal lines. HD~41641 was found to have $T_\mathrm{eff}$ = 7200 $\pm$ 80~K, \ensuremath{\log g} = 3.5 $\pm$ 0.3 dex, \ensuremath{{\upsilon}\sin i}~=~30~$\pm$~2~$\mathrm{km\,s}^{-1}$, $\xi$~=~1.1~$\pm$~0.3~$\mathrm{km\,s}^{-1}$, and a metallicity of -0.19~$\pm$~0.08~dex (see Table \ref{params}). The included uncertainties are those computed from the $\chi^2$ minimization, using a 1$\sigma$ confidence level. Figure~\ref{GSSPfit} shows the best-fitting synthetic spectrum (dashed black line) to the average spectrum (solid gray line) for two different spectral regions. 

\begin{table}[tb] 
\caption{Stellar parameters obtained with the GSSP code \citep{GSSPnew} and estimated values of  the mass and radius of HD~41641.}
\begin{center}
\begin{tabular}{rl}
\hline\hline
\rule[0mm]{0mm}{4mm}
\textit{RV} & 29.3 $\pm$ 0.3 $\mathrm{km\,s}^{-1}$\\
\rule[0mm]{0mm}{4mm}
$T_\mathrm{eff}$ & 7200 $\pm$ 80 K\\
\rule[0mm]{0mm}{4mm}
\ensuremath{\log g} & 3.5 $\pm$ 0.3 dex\\
\rule[0mm]{0mm}{4mm}
\ensuremath{{\upsilon}\sin i} & 30 $\pm$ 2 $\mathrm{km\,s}^{-1}$\\ 
\rule[0mm]{0mm}{4mm}
[\textit{M}/\textit{H}] & -0.19 $\pm$ 0.08 dex\\
\rule[0mm]{0mm}{4mm}
$\xi$ & 1.1 $\pm$ 0.3 $\mathrm{km\,s}^{-1}$\\
\rule[0mm]{0mm}{4mm}
\textit{M}$_*$ & 2.3$^{+0.7}_{-0.5}$~\textit{M}$_{\sun}$\\
\rule[0mm]{0mm}{4mm}
\textit{R}$_*$ & 4.5$^{+2.7}_{-1.7}$~\textit{R}$_{\sun}$\\
\hline\hline
\end{tabular}
\end{center}
\label{params}
\end{table}

\begin{figure}[tb]
\centering
\includegraphics[width=0.49\textwidth]{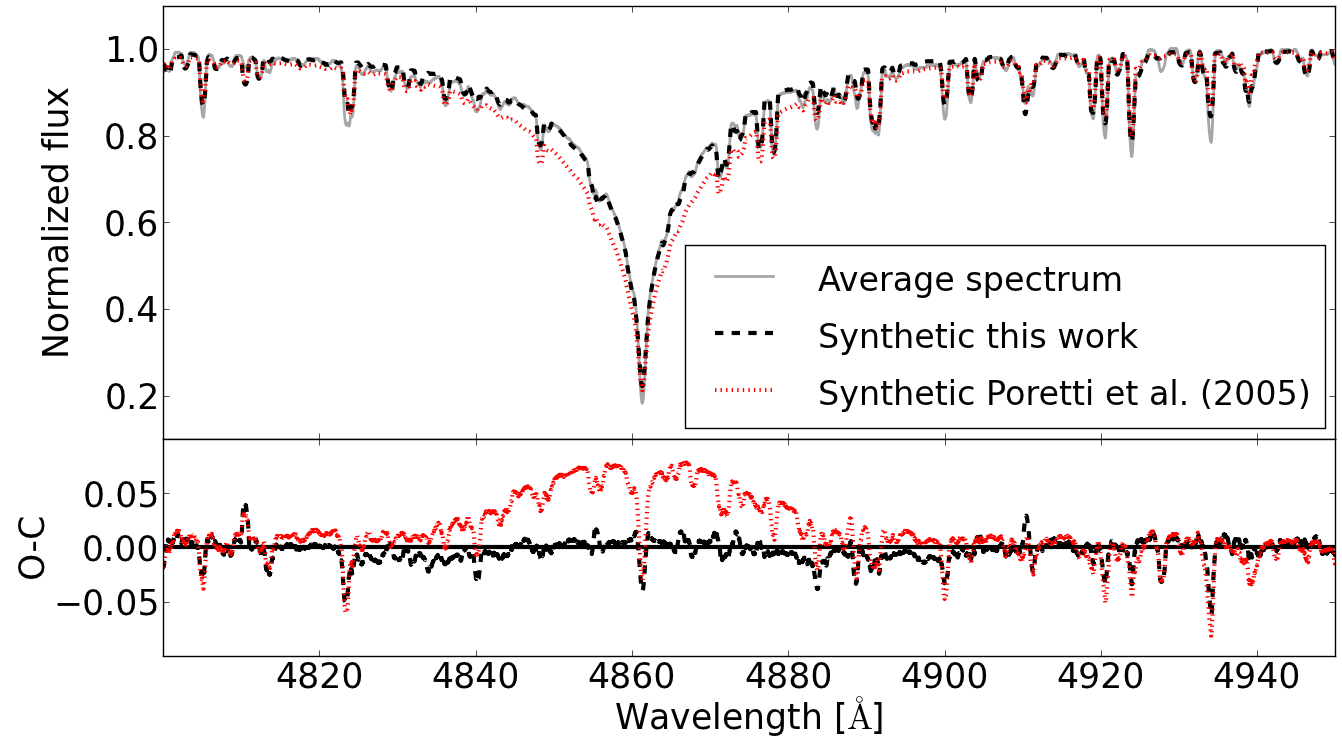}
\caption{\label{compPar} Comparison of our results, which are plotted with a dashed black line, with the results obtained by \cite{ThisStar}, which are plotted with a dotted red line.}
\end{figure}

\begin{figure}[htb]
\centering
\includegraphics[width=0.49\textwidth,clip]{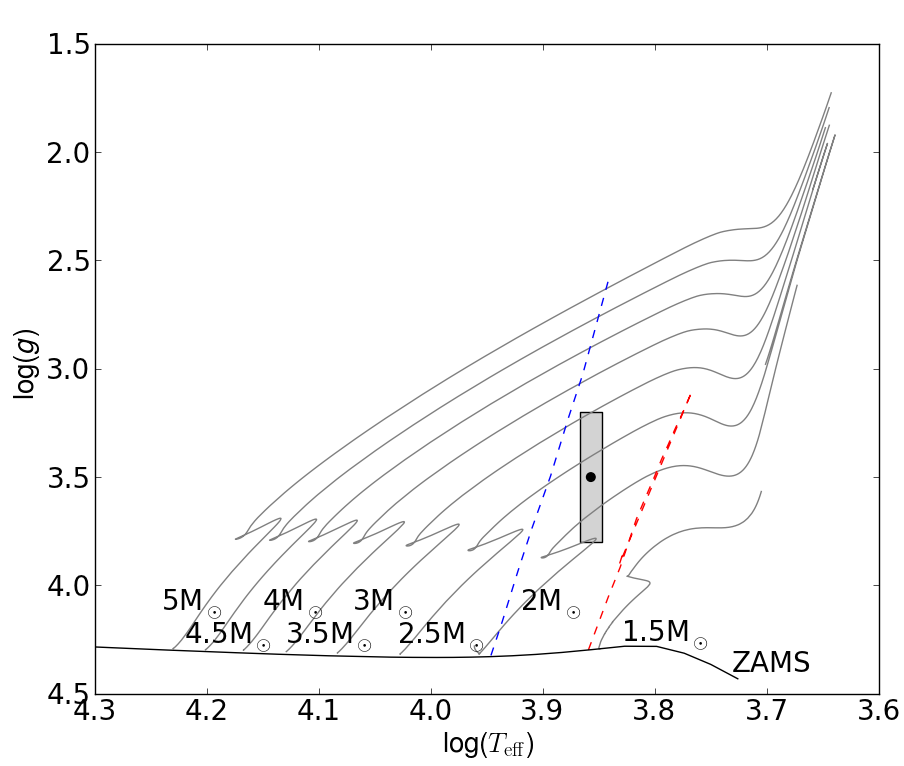}
\caption{\label{tracks} Position of HD 41641 (black dot) in a (\ensuremath{\log g}-log$T_\mathrm{eff}$) diagram. The shaded box corresponds to the error bars obtained from the $\chi^2$ minimization. The ZAMS (black solid line), theoretical blue edge (blue dashed line), and empirical red edge (red dashed line) of the classical $\delta$~Sct instability strip were taken from \cite{BregerPamyatnykh}. Post-main-sequence evolutionary tracks (gray solid lines) are from \cite{tracks}.}
\end{figure}

The values of \ensuremath{{\upsilon}\sin i}~$\sim$~29~$\mathrm{km\,s}^{-1}$ and $[M/H]$~=~-0.2~$\pm$~0.2~dex reported in the discovery paper \citep{ThisStar} are in perfect agreement with ours, and \ensuremath{\log g}~=~3.9~$\pm$~0.2~dex agrees within $1\sigma$, but their $T_\mathrm{eff}$~=~7700~$\pm$~200~K is slightly different. \cite{ThisStar} used Str\"{o}mgren $uvby\beta$ photometry, while we determined our values from spectroscopy. In order to actually compare both results, a synthetic spectrum was generated with the GSSP code via the four parameters obtained by \cite{ThisStar} and setting $\xi$ as free. From the fit of the $H\beta$ line (Fig.~\ref{compPar}), it is seen that the parameters derived from our work give a better result.

The estimated values of $T_\mathrm{eff}$ and \ensuremath{\log g} allow us to place the target into a (\ensuremath{\log g}-log$T_\mathrm{eff}$) diagram. We accomplish this in Fig.~\ref{tracks}, where the solid circle represents HD 41641, and the shaded box corresponds to the error bars obtained from the $\chi^2$ minimization. The evolutionary tracks were taken from \cite{tracks}. The theoretical blue edge of the classical $\delta$~Sct instability strip and the empirical red edge are also plotted, as well as the zero-age main sequence (ZAMS) \citep{BregerPamyatnykh}. The derived atmospheric parameters place the star out of the main sequence, indicating its advanced evolution.

From this, we estimate the mass of HD~41641 to be \textit{M}$_*$~=~2.3$^{+0.7}_{-0.5}$~\textit{M}$_{\sun}$. Finally, the radius of the star could be estimated to be \textit{R}$_*$~=~4.5$^{+2.7}_{-1.7}$~\textit{R}$_{\sun}$, from the derived \ensuremath{\log g} and estimated mass range.

\subsection{Analysis of the chemical abundances}\label{ssec:chem}
An intensive study of the individual chemical abundances was also performed with the previously described GSSP code by fitting the average spectrum. Each element was treated individually, depending on the number of lines that it presents on the spectra. For example, for line-rich elements such as iron, a wide wavelength range was used. Additionally, the other four stellar parameters (i.e.,$T_\mathrm{eff}$, \ensuremath{\log g}, \ensuremath{{\upsilon}\sin i}, and $\xi$) were set free, as iron has an important influence on them and fixing them would have imposed constraints on its abundance. On the other hand, for the analysis of elements with less spectral lines, the stellar parameters were fixed, and only small wavelength ranges, where those lines are known to be present, were fitted. The obtained abundances were compared with solar values \citep{solarab} and the results are summarized in Table \ref{ChemAbund}. Only those elements whose abundance could be properly constrained are included in the table. The presented uncertainties were derived from the 1$\sigma$ confidence level of the $\chi^2$ minimization procedure. 

Several chemical elements, indicated in Table \ref{ChemAbund} with an asterisk, showed different abundances than expected for a star of the derived global metallicity. For some of these abundances, the deviation is small taking the uncertainties into account (e.g., Mg, Ca, or Ni), but others present quite divergent values. Mn, Co, Y, Ba, and Nd show considerable overabundance, and Sc, Cu, and Zn were determined to be underabundant. For some other elements (e.g., V and Zr), it was impossible to reduce the uncertainties, even when working with different spectral ranges. The overabundances of V, Mn, and Co may be partially related with the fact that we did not take the hyperfine structure into account. However they are not expected to be larger than 0.2~dex.

\begin{figure}[tb]
\centering
\includegraphics[width=0.50\textwidth]{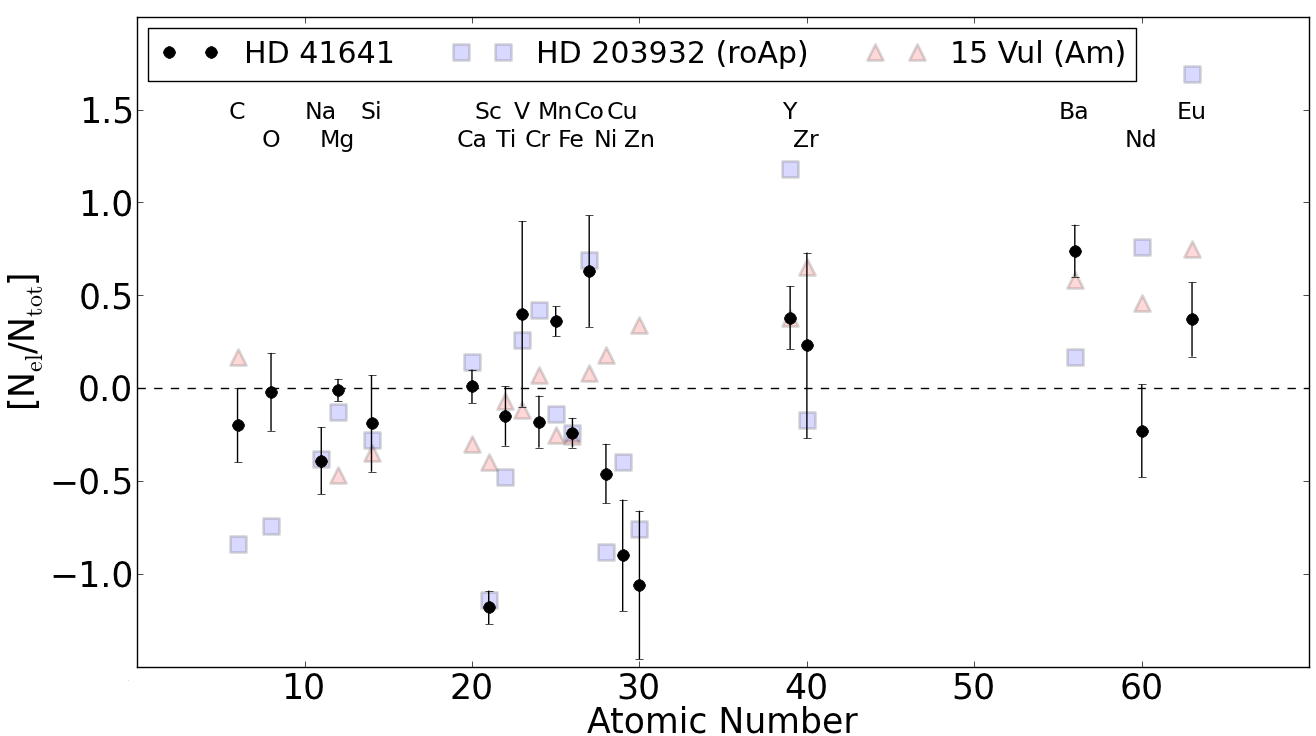}
\caption{Individual chemical abundances of HD~41641 (black dots) compared with those of HD~203932 (blue squares), a roAp star \citep{roAp_plot}, and 15~Vul (red triangles), an Am star \citep{microt}.}
\label{abund} 
\end{figure}

The observed abundance pattern shows characteristics of an object in between classical Am and mild Ap stars. Figure \ref{abund} shows a comparison of the individual chemical abundances of HD~41641 with those of HD~203932, a roAp star with a small magnetic field \citep{roAp_plot}, and 15~Vul, an Am star \citep{microt}. It can be observed that the Fe-peak behavior, as well as the Na, Mg, and Si abundances, are much closer to an Ap than to an Am star. It is remarkable that we find a difference in the Nd abundance, when we analyze the \ion{Nd}{II} lines (4706.54\text{\AA}, 4959.119\text{\AA}, 5076.560\text{\AA}, 5092.79\text{\AA}, and 6293.16\text{\AA}) and the \ion{Nd}{III} lines (5050.695\text{\AA}, 5102.428\text{\AA}, 5203.924\text{\AA}, and 5294.113\text{\AA}). This effect, called Nd-anomaly, is typically observed in cool Ap stars, and mostly in the roAp stars \citep{Ryabchikova04}. The roAp stars with the longest pulsation periods are usually more evolved than the short-period roAp stars and the Nd-anomaly is typically the same as in HD~41641 \citep{Alentiev12}. The derived atmospheric parameters show that HD~41641 is an evolved star. A few evolved Ap stars are known to have low rare-earth surface abundances and a weak but detectable magnetic field \citep{Titarenko12}. The determined abundances allow us to suggest that HD~41641 is an Ap star even when it shows pure $\delta$~Sct-type pulsations (see Sect. \ref{sec:freqana}).

Besides the chemistry, there are two more arguments for HD~41641 to be an Ap star rather than an Am star. One is the low microturbulent velocity. For Am stars, this parameter varies between 3 $\mathrm{km\,s}^{-1}$ and 5 $\mathrm{km\,s}^{-1}$ \citep{microt}, while Ap stars are expected to have lower values, from 0~$\mathrm{km\,s}^{-1}$ to 2~$\mathrm{km\,s}^{-1}$. The second argument is that about 90\% of the known Am stars are in binary systems \citep{Am_bin}. HD~41641 does not have an orbital companion (see Sect. \ref{ssec:lowfreq}), which is another indication of the fact that its identification as an Ap star would be more likely. 

\begin{table}[bt]
\caption{Individual chemical abundances of HD~41641 and comparison with solar values \citep{solarab}.}
\begin{center}
\small
\begin{tabular}{cccc}
\hline\hline
\rule[0mm]{0mm}{4mm}
 \textbf{Chemical} & \textbf{Abundance}& \textbf{Solar}  & \textbf{Difference}\\
\rule[0mm]{0mm}{3mm}
\textbf{Element} & \textbf{log(\textit{N}$_{\text{el}}$/\textit{N}$_{\text{tot}}$)}& \textbf{abundance} & \textbf{to solar}\\
\hline 
\rule[0mm]{0mm}{4mm}
\textbf{C} & -3.81 $\pm$ 0.20 & -3.61 & -0.2 $\pm$ 0.20 \\
\rule[0mm]{0mm}{4mm}
\textbf{O} & -3.37 $\pm$ 0.21 & -3.35 & -0.02 $\pm$ 0.21 \\
\rule[0mm]{0mm}{4mm}
\textbf{Na} & -6.09 $\pm$ 0.18 & -5.70 & -0.39 $\pm$ 0.18\\
\rule[0mm]{0mm}{4mm}
\textbf{Mg*} & -4.52 $\pm$ 0.06 & -4.51 & -0.01 $\pm$ 0.06\\
\rule[0mm]{0mm}{4mm}
\textbf{Si} & -4.72 $\pm$ 0.26 & -4.53 & -0.19 $\pm$ 0.26\\
\rule[0mm]{0mm}{4mm}
\textbf{Ca*} & -5.69 $\pm$ 0.09 & -5.70 & +0.01 $\pm$ 0.09\\
\rule[0mm]{0mm}{4mm}
\textbf{Sc*}& -10.17 $\pm$ 0.09 & -8.99 & -1.18 $\pm$ 0.09\\
\rule[0mm]{0mm}{4mm}
\textbf{Ti} & -7.24 $\pm$ 0.16 & -7.09 & -0.15 $\pm$ 0.16\\
\rule[0mm]{0mm}{4mm}
\textbf{V} & -7.71$^{+0.36}_{-1.00}$ & -8.11 & +0.39$^{+0.36}_{-1.00}$\\
\rule[0mm]{0mm}{4mm}
\textbf{Cr} & -6.58 $\pm$ 0.14 & -6.40 & -0.18 $\pm$ 0.14\\
\rule[0mm]{0mm}{4mm}
\textbf{Mn*} & -6.25 $\pm$ 0.08 & -6.61 & +0.36 $\pm$ 0.08\\
\rule[0mm]{0mm}{4mm}
\textbf{Fe} & -4.78 $\pm$ 0.08 & -4.54 & -0.24 $\pm$ 0.08\\
\rule[0mm]{0mm}{4mm}
\textbf{Co*} & -6.39$^{+0.29}_{-0.45}$ & -7.02 & +0.63$^{+0.29}_{-0.45}$ \\
\rule[0mm]{0mm}{4mm}
\textbf{Ni*} & -6.28 $\pm$ 0.16  & -5.82 & -0.46 $\pm$ 0.16\\
\rule[0mm]{0mm}{4mm}
\textbf{Cu*} & -8.75 $\pm$ 0.30 & -7.85 & -0.90 $\pm$ 0.30 \\
\rule[0mm]{0mm}{4mm}
\textbf{Zn*} & -8.54$^{+0.31}_{-0.69}$ & -7.48 & -1.06$^{+0.31}_{-0.69}$ \\
\rule[0mm]{0mm}{4mm}
\textbf{Y*} & -9.45 $\pm$ 0.17 & -9.83 & +0.38 $\pm$ 0.17 \\
\rule[0mm]{0mm}{4mm}
\textbf{Zr} & -9.23 $\pm$ 0.50 & -9.46 & +0.23 $\pm$ 0.50\\
\rule[0mm]{0mm}{4mm}
\textbf{Ba*} & -9.12 $\pm$ 0.14 & -9.86 & +0.74 $\pm$ 0.14 \\
\rule[0mm]{0mm}{4mm}
\textbf{\ion{Nd}{II}} & -10.85 $\pm$ 0.26 & -10.62 & -0.23 $\pm$ 0.26 \\
\rule[0mm]{0mm}{4mm}
\textbf{\ion{Nd}{III}*} & -10.04 $\pm$ 0.24 & -10.62 & +0.58 $\pm$ 0.24 \\
\rule[0mm]{0mm}{4mm}
\textbf{\ion{Eu}{II}*} & -11.15 $\pm$ 0.20 & -11.52 & +0.37 $\pm$ 0.20 \\
\hline\hline
\end{tabular}
\end{center}
\label{ChemAbund}
\end{table}

\section{Frequency analysis}\label{sec:freqana}

\subsection{Frequencies from photometry in the $\delta$ Sct range}\label{ssec:CoRoTfreq}

The software package Period04 \citep{Period04}, which is based on classical Fourier analysis techniques and least-squares algorithms, was used to perform the frequency analysis on the photometric data. A frequency was considered to be significant when its S/N was higher than 4.0 \citep{Breger93}.

\begin{figure*}[tb]
\centering
\includegraphics[width=0.45\textwidth]{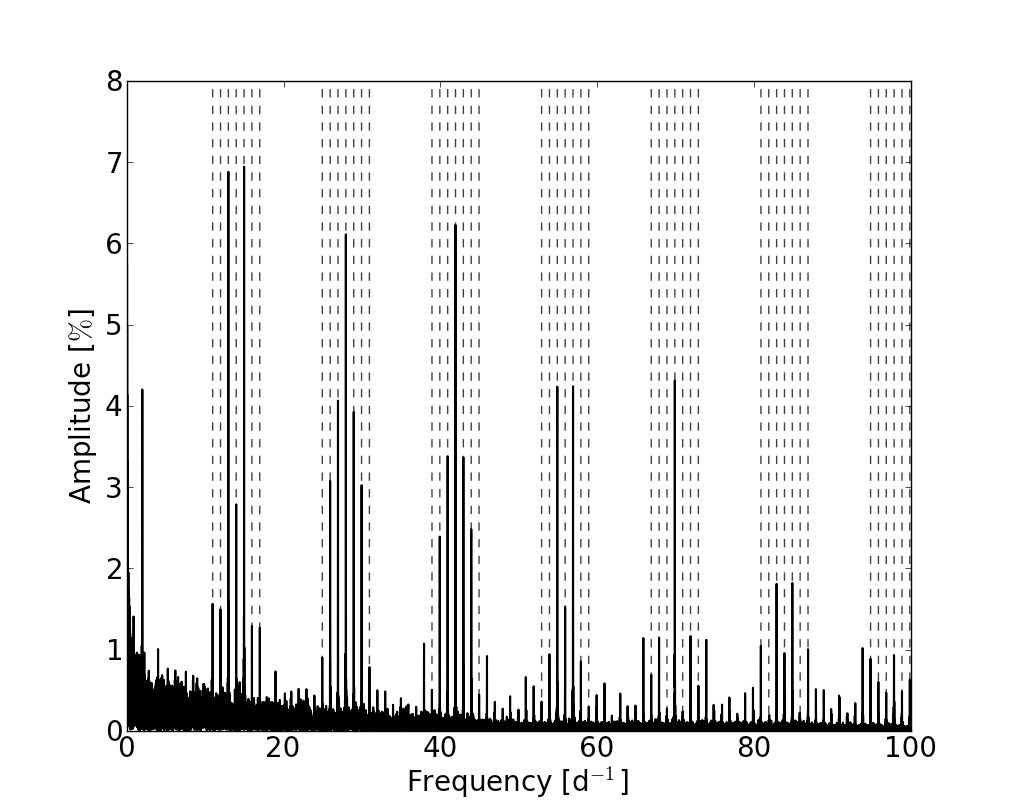}
\includegraphics[width=0.45\textwidth]{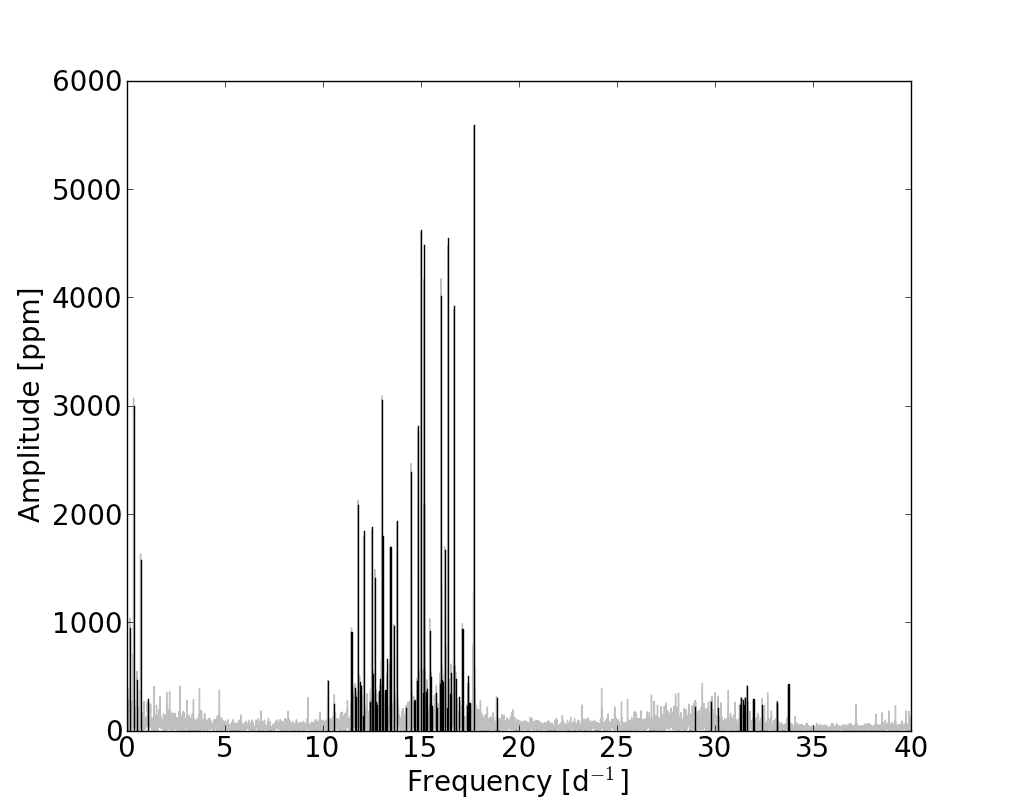}
\caption{Computed spectral window of the photometric data set, where the frequencies related to the observational window are indicated with dashed lines (left), and the range (<40 d$^{-1}$) of the Fourier spectrum, where the significant frequencies of the target were detected (right).}
\label{fou} 
\end{figure*}

The left panel of Fig.~\ref{fou} shows the spectral window of the CoRoT light curve. As the amplitude of the highest alias peak is less than 8$\%$, the plot presents a zoom into the Y-axis. Most of the peaks correspond to the orbital frequency of the satellite (i.e., 13.972 d$^{-1}$) and its aliases \citep{CoRoT}. These are indicated with dashed vertical lines. There is a peak with remarkable amplitude at $\sim$2 d$^{-1}$, which is related to the CoRoT passage through the South Atlantic Anomaly, which occurs twice per sidereal day.

The right panel of Fig.~\ref{fou} shows the region of the Fourier spectrum where intrinsic stellar frequency peaks could be observed (<40 d$^{-1}$). A total of 90 significant frequencies was found. Most of the peaks appear over the threshold of a few d$^{-1}$ and are within the expected range to be caused by the excitation of pressure modes. Additionally, there are several peaks at lower frequencies, which cannot be independent p-mode frequencies and are discussed later (see Sect.~\ref{ssec:lowfreq}).

Table \ref{freqs} lists all the detected frequencies and their amplitudes in the Fourier spectrum. The digits in parentheses are the errors. These were computed by multiplying the formal errors described by \cite{errors} with a factor 10. \cite{errors_Degroote} showed, using CoRoT data, that the traditional assumptions of uncorrelated flux points and white Gaussian noise are not realistic, and that the mentioned formal errors are too optimistic. Additionally, a subset of our residual light curve can be observed in the lower panel of Fig.~\ref{LCzoom}. It is clear that all the signal has not been prewhitened and there is still some correlation between the points. We analyzed this correlation, as described in \cite{errors_correlation} and decided to adopt error bars of an order of magnitude wider than the formal error bars. Finally,  columns 4, 8, and 12 indicate possible combination frequencies. All the frequencies close or within the frequency resolution of an already listed component were ignored to avoid spurious frequencies due to the high peak density in the power spectrum.

Inside the error bars, every peak present in the rightmost group (>25 d$^{-1}$) could be explained as combination of other frequencies. This allows us to exclude this range of frequencies as independent frequencies. This is in full agreement with a recent study by \cite{Kurtz15} about the presence of combination frequencies in the Fourier spectrum of several types of pulsators. Hence, the oscillation frequency range of HD~41641 is reduced to the range between 10 d$^{-1}$ and 20 d$^{-1}$. These values are far from the expected frequencies of a roAp star and allow us to confirm the classification of HD~41641 as a pure $\delta$~Sct pulsator.

\begin{table*}[tb]
\caption{Results from the frequency analysis of the photometric data set of HD 41641.} 
\label{freqs}
\begin{center}
\begin{scriptsize}
\begin{tabular}{cccc|cccc|cccc}
\hline
\hline
\rule[0mm]{0mm}{3mm}
 & \textbf{Freq} & \textbf{Ampl $\pm$ 50} &  &  & \textbf{Freq} & \textbf{Ampl $\pm$ 50} &  &  & \textbf{Freq} & \textbf{Ampl $\pm$ 50} & \\
\rule[0mm]{0mm}{2mm}
\textbf{F\#} & \textbf{[d$^{-1}$]} & \textbf{[ppm]} & \textbf{Relations} & \textbf{F\#} & \textbf{[d$^{-1}$]} & \textbf{[ppm]} & \textbf{Relations} & \textbf{F\#} & \textbf{[d$^{-1}$]} & \textbf{[ppm]} & \textbf{Relations}\\
\hline
\rule[0mm]{0mm}{3mm}
F1 & 17.70259(6) & 5590 &  & F31 & 15.5191(6) & 500  &  & F61 & 11.610(1)  & 310  & \\
F2 & 15.01565(7) & 4630 &  & F32 & 17.4212(6) & 510  &  & F62 & 1.069(1)   & 290  & 6 * F24 \\
F3 & 16.38174(7) & 4550 &  & F33 & 12.5586(6) & 520  &  & F63 & 31.957(1)  & 290  & F7 + F39 + F81 \\
F4 & 15.16579(7) & 4480 &  & F34 & 16.7766(6) & 480  &  & F64 & 31.332(1)  & 300  & F27 + F36 + F41 \\
F5 & 16.03780(8) & 4010 &  & F35 & 10.2663(7) & 460  &  & F65 & 12.519(1)  & 300  & \\
F6 & 16.68883(8) & 3920 &  & F36 & 0.5342(7)  & 470  & 3 * F24 & F66 & 31.428(1)  & 290 & F7 + F51 + F53\\
F7 & 0.3560(1)   & 3000 & 2 * F24 & F37 & 11.8797(7) & 450  &  & F67 & 14.691(1)  & 280  & \\
F8 & 13.0061(1)  & 3050 &  & F38 & 14.8187(7) & 460  & F30 - F7 & F68 & 12.715(1)  & 260  & \\
F9 & 14.8545(1)  & 2820 &  & F39 & 16.0538(7) & 470  & F31 + F36 & F69 & 12.428(1)  & 270  & \\
F10 & 14.5109(1) & 2390 &  & F40 & 16.1197(7) & 450  &  & F70 & 29.770(1)  & 270  & F5 + F36 + F49 \\
F11 & 15.1526(1) & 2090 &  & F41 & 17.3914(7) & 440  &  & F71 & 14.639(1)  & 270  & \\
F12 & 11.7942(1) & 2090 &  & F42 & 33.7405(7) & 440  & F5 + F7 + F82 & F72 & 12.396(1)  & 260  & \\
F13 & 13.7751(2) & 1940 &  & F43 & 11.9434(8) & 410  &  & F73 & 15.484(1)  & 250 & \\
F14 & 12.49191(2) & 1880 &  & F44 & 15.9668(7) & 440  & F41 - 4*F7 & F74 & 17.480(1)  & 250  & \\
F15 & 12.0821(2) & 1850 &  & F45 & 31.6296(8) & 410  & F36 + F81 + F81 & F75 & 10.590(1)  & 240  & F47 - F19 - F24 \\
F16 & 13.0858(2) & 1800 &  & F46 & 11.6353(8) & 400  &  & F76 & 32.406(1)  & 240  & F3 + F19 + F48 \\
F17 & 13.4505(2) & 1700 &  & F47 & 11.4796(8) & 410  &  & F77 & 31.249(1)  & 240  & F3 + F7 + F10 \\
F18 & 16.2143(2) & 1670 &  & F48 & 15.3114(8) & 380  &  & F78 & 12.745(1)  & 240  & \\
F19 & 0.7124(2)  & 1570 & 4 * F24 & F49 & 13.19819(8) & 380  &  & F79 & 33.174(1)  & 250  & F23 + F31 + F36 \\
F20 & 12.6527(2) & 1410 &  & F50 & 0.01396(8) & 390  &  & F80 & 31.487(1)  & 240  & F24 + F28 + F71 \\
F21 & 11.4657(3) & 910  &  & F51 & 15.2804(9) & 360  &  & F81 & 15.548(1)  & 230  & F2 + F36 \\
F22 & 13.6165(3) & 970  &  & F52 & 12.8641(8) & 370  &  & F82 & 17.347(1)  & 230  & F1 - F7 \\
F23 & 17.1205(3) & 940  &  & F53 & 15.7921(9) & 340  &  & F83 & 28.978(1)  & 220  & F7 + F15 + F29 \\
F24 & 0.17756(3) & 950  & f$_{rot}$ & F54 & 15.1366(9) & 350  &  & F84 & 12.9364(6) & 470  & \\
F25 & 15.4545(3) & 920  &  & F55 & 16.5018(9) & 340  & F44 + F36 & F85 & 12.9365(7) & 420  & \\
F26 & 13.2722(5) & 660  &  & F56 & 12.5416(9) & 350  &  & F86 & 14.261(1)  & 210  & \\
F27 & 13.4060(5) & 640  &  & F57 & 31.534(1)  & 300  & F9 + F19 + F44 & F87 & 16.330(1)  & 210  & \\
F28 & 16.6709(5) & 600  &  & F58 & 16.950(1)  & 310  &  & F88 & 30.159(1)  & 210  & F6 + F36 + F84 \\
F29 & 16.5397(5) & 530  &  & F59 & 33.157(1)  & 260  & F34 + F48 + F62 & F89 & 15.811(1)  & 210  & F25 + F7\\
F30 & 15.1752(5) & 570  &  & F60 & 18.859(1)  & 310  &  & F90 & 12.068(2)  & 140  & \\
\hline\hline
\end{tabular}
\end{scriptsize}
\end{center}
\tablefoot{Frequencies, amplitudes, and detected linear combinations between frequencies. The last-digit errors of the frequencies are given in parentheses. They were computed by multiplying the formal errors \citep{errors} by ten to take the correlated nature of the data into account. The uncertainties on the amplitudes are also the formal errors multiplied by a factor 10.}
\end{table*}

A comparison of the oscillation periods with theoretical Q-values may help us to interpret the Fourier spectrum of the star \citep{BregerQ}. The values of the pulsational constant, $Q_i$, associated with each detected p-mode frequency in a $\delta\,$Sct star can be estimated using equation \ref{eqQ} \citep{Breger93} as follows:
\begin{equation}\label{eqQ}
\log Q_i = -6.456 + \log P_i + 0.5\log g + 0.1M_{\mathrm{bol}} + \log T_{\mathrm{eff}},
\end{equation}
\noindent where $P_i$ is the period that corresponds to each oscillation frequency, \ensuremath{\log g} and $T_\mathrm{eff}$ were determined in Sect. \ref{sec:param}, and $M_{\mathrm{bol}}$ can be estimated from \textit{R}$_*$ and $T_\mathrm{eff}$.

The pulsational constants obtained for the oscillation frequencies of HD~41641 are smaller than 0.016. Although the uncertainties of our parameters are large, if we compare the obtained values with those listed in Table II of \cite{Qdelta}, we can conclude that all the detected frequencies are overtone modes. On the other hand, for cool $\delta$~Sct stars, the lowest pulsation frequency is expected to be close to the radial fundamental mode frequency. If we use the theoretical Q-value listed in the mentioned table, the fundamental mode frequency of HD~41641 should be detected at about 4.83 d$^{-1}$. We do not detect any pulsation with a frequency close to this value.

Rotationally split frequencies have been detected in a few slowly-rotating $\delta$~Sct stars (e.g., \citealt{Kurtz14rot}; \citealt{Saio15rot}). We searched for rotational splitting, which should be on the order of the average angular velocity for high-order or high-degree acoustic modes \citep{Cunha07}, but do not find any clear signature of frequency splitting.

\subsection{Frequencies from spectroscopy in the $\delta$ Sct range}\label{ssec:HARPSfreq}

\begin{figure*}[t]
\centering
\includegraphics[width=0.33\textwidth]{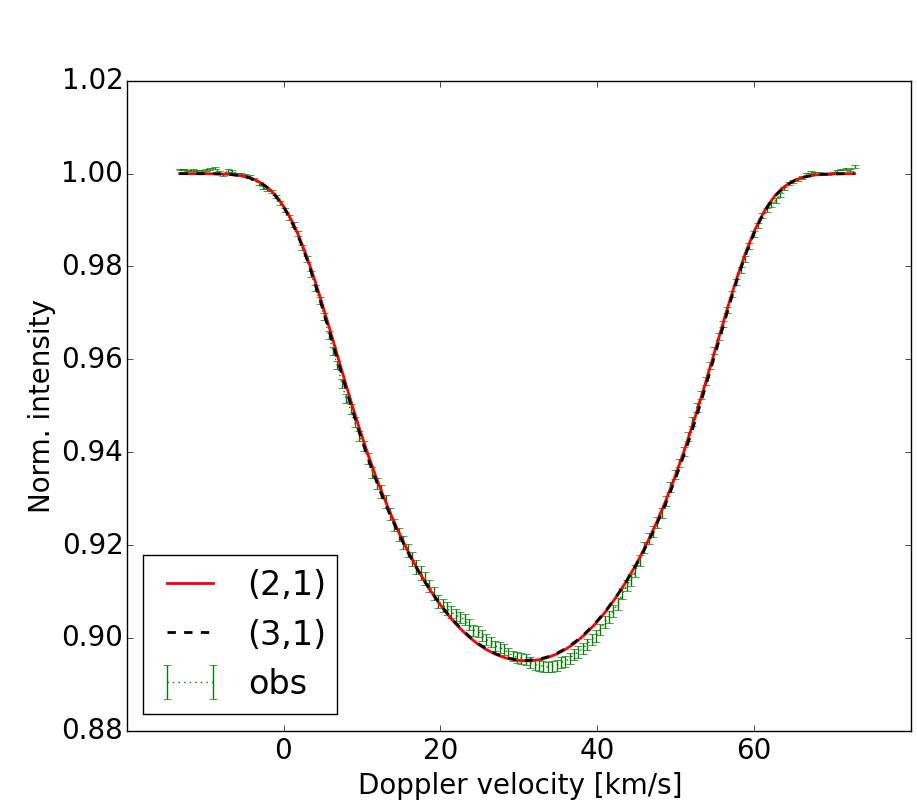}
\includegraphics[width=0.33\textwidth]{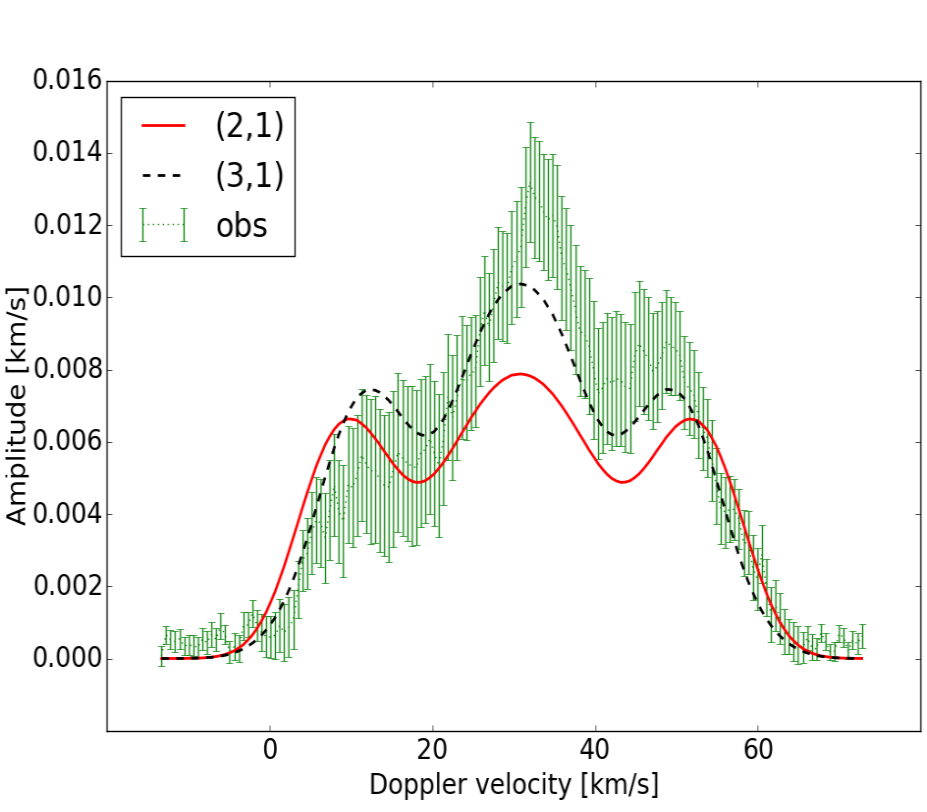}
\includegraphics[width=0.33\textwidth]{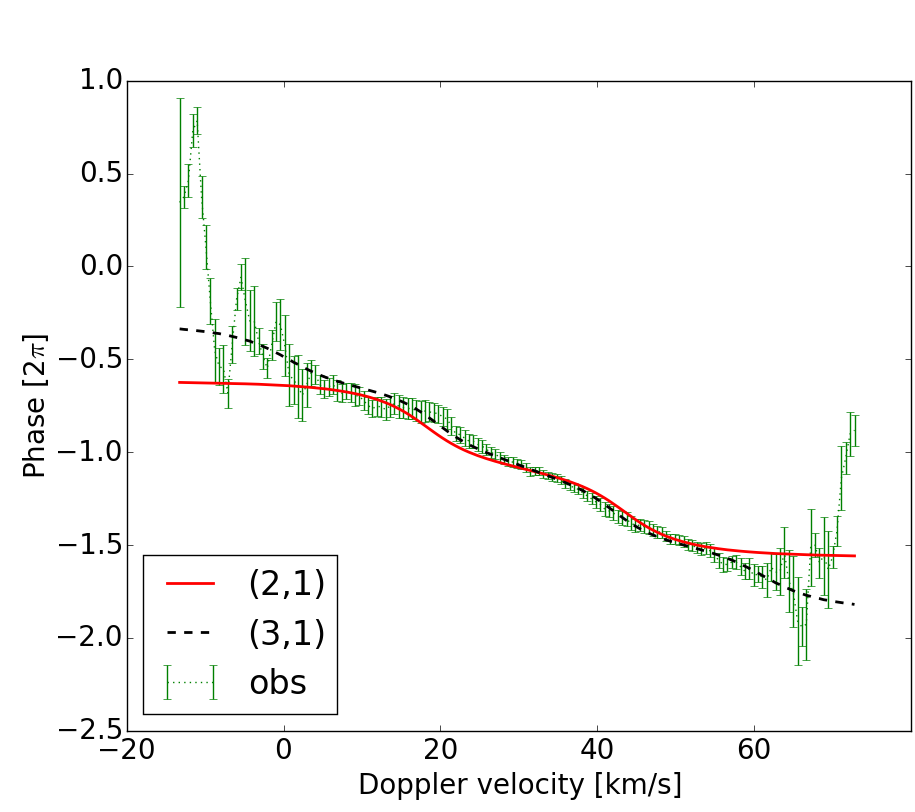}
\caption{Identification of the individual mode geometry for F1 studying line 1. The plots correspond to the normalized intensity of the studied spectral line (left), amplitude of the variation (middle), and phase (right). The observations are plotted with a green dotted line and the error bars are included in green as well. The two most suitable identifications are plotted with a dashed black line for (\textit{l},\textit{m})~=~(3,1), which was obtained with a $\chi^2$ of 1.88, and a solid red line for (\textit{l},\textit{m})~=~(2,1), which was obtained with a $\chi^2$ of 2.43.}
\label{F1} 
\end{figure*}

Timely variations of several spectral lines were studied with the goal of identifying the excited modes in HD~41641. We used the software tool Frequency
Analysis and Mode Identification for Asteroseismology (FAMIAS; \citealt{FAMIAS}). It is based on Fourier analysis techniques and least-squares fitting algorithms to detect pulsation frequencies and identify the associated pulsation mode geometries. We applied the moment method \citep{momentmethod} and the Fourier parameter fit (FPF) method \citep{Zima06_method} and compared the results. 

A careful selection of the used lines is crucial. We looked for unblended, deep, and sharp lines with narrow wings, in which the line profile variation was detectable across the whole profile and not only at the line center. Since HD~41641 has a relatively low projected rotational velocity (see Table \ref{params}), lines were narrow in general. We selected six iron lines, three of which were of \ion{Fe}{I} ($\lambda_0$ = 5367.467\text{\AA}, 5383.369\text{\AA}, and 5393.168\text{\AA}), while the other three were of \ion{Fe}{II} ($\lambda_0$ = 5362.869\text{\AA}, 4508.288\text{\AA}, and 4549.479\text{\AA}). Additionally, three other metal lines (\ion{Cr}{II}: $\lambda_0$~=~5237.32\text{\AA}, \ion{Ti}{II}: $\lambda_0$~=~4779.985\text{\AA}, and \ion{Ni}{I}: $\lambda_0$~=~4786.531\text{\AA}) were selected. All of these were carefully normalized by fitting the continuum with a linear spline function with the code described in Sect. \ref{ssec:HARPS}. It is important to notice that the exposure times of the spectra cover about one-quarter of the pulsation cycle for the oscillations with the highest frequencies. This could introduce pulsational broadening in the spectral line affecting our \ensuremath{{\upsilon}\sin i} determination, which may have been overestimated (for confusion between rotational and pulsational broadening, see also \citealt{Aerts2014}).

For the nine selected lines, the Fourier spectrum of the data was computed to search for periodicities. In the case of the FPF method, we used the Fourier transform of the line profile, while we worked with the Fourier transform of each n$^{th}$ moment independently for the moment method. The significance level required to accept a frequency was chosen to be S/N $>$ 4.0. Table \ref{freqFAM} lists the p-mode oscillation frequencies, which were found using any of the methods in the frequency analysis of each spectral line. Most of these frequencies correspond to frequencies found in the photometric data. Only one frequency within the p-mode range of HD~41641 is not part of the photometric frequency list. This frequency is f = 14.448 $\pm$ 0.005 d$^{-1}$ and was detected in two of the nine spectral lines.

Identification of the individual mode geometries was attempted with the two mentioned methods, but the dominance of the low frequencies was so strong that it was not successful. With the goal of reducing the influence of the low frequencies, the spectra were phase-folded using each one of the p-mode frequencies detected. Then, using phase bins of 5$\%$ width of the phase, the spectra of each bin were averaged. In this way, the variations related to any other frequency were smeared out, and the identification of the individual mode geometries could be performed with the FPF method, treating each frequency individually and imposing the value found in photometry (Table \ref{freqs}).

\begin{table}[tb] 
\caption{Selected lines to perform the spectrosopic frequency analysis and obtained oscillation frequencies.}
\begin{center}
\begin{tabular}{cccl}
\hline\hline
\rule[0mm]{0mm}{4mm}
 & \textbf{Spectral} & \textbf{Central wavelength} & \textbf{Oscillation} \\
\rule[0mm]{0mm}{3mm}
\# & \textbf{line} & \textbf{$\lambda_0$ [\text{\AA}]} & \textbf{frequencies} \\
\hline
\rule[0mm]{0mm}{3mm}
1 & \ion{Fe}{I} & 5367.467 & F1, F5, F15, F40\\
\rule[0mm]{0mm}{3mm}
2 & \ion{Fe}{I} & 5383.369 & F1, F15\\
\rule[0mm]{0mm}{3mm}
3 & \ion{Fe}{I} & 5393.168 & F5, F6, F40, f\\
\rule[0mm]{0mm}{3mm}
4 & \ion{Fe}{II} & 5362.869 & F1, F40\\
\rule[0mm]{0mm}{3mm}
5 & \ion{Fe}{II} & 4508.288 & F1, F5, F6, F40, f\\
\rule[0mm]{0mm}{3mm}
6 & \ion{Fe}{II} & 4549.479 & F6, F40\\
\rule[0mm]{0mm}{3mm}
7 & \ion{Cr}{II} & 5237.320 & F1, F40\\
\rule[0mm]{0mm}{3mm}
8 & \ion{Ti}{II} & 4779.985 & F1, F5, F6, F40\\
\rule[0mm]{0mm}{3mm}
9 & \ion{Ni}{I} & 4786.531 & F1, F6, F40\\
\hline\hline
\end{tabular}
\end{center}
\label{freqFAM}
\tablefoot{Only the pulsation-related frequencies (between 10 d$^{-1}$ and 20 d$^{-1}$) are included, and the notation is consistent with that used during the photometric analysis. f = 14.448 $\pm$ 0.005 d$^{-1}$ is the only frequency detected only in the spectroscopic data set.}
\end{table}

We could only  put  constraints on the surface geometry in terms of the (\textit{l},\textit{m}) quantum numbers for the oscillation mode related to F1. The analysis suggests that the frequency is associated with a nonradial prograde mode with \textit{m}~=~+1. The mode with (\textit{l},\textit{m})~=~(3,1) gives the lowest $\chi^2$ value for the analysis of all the spectral lines in which F1 was detected. Figure~\ref{F1} shows the fit of the model (dashed black line) to the observations (dotted green line) for line 1. The left panel corresponds to the normalized intensity of the spectral line, the middle panel corresponds to the amplitude, and the right panel corresponds to the phase. The error bars related to the observations are included. The inclination angle obtained for this most favorable solution is \textit{i}~=~45\degr $\pm$ 5\degr and the $\chi^2$ value is 1.88. Other possible identifications are (\textit{l},\textit{m})~=~(2,1) with \textit{i}~=~50\degr $\pm$ 5\degr and $\chi^2$ = 2.43, which is also included in Fig.~\ref{F1} with a continuous red line, (\textit{l},\textit{m}) = (3,3) with $\chi^2$ = 2.65, and (\textit{l},\textit{m}) = (1,1) with $\chi^2$ = 3.22.

The observed amplitude of an oscillation mode strongly depends on the orientation of the stellar rotation axis and its inclination with respect to the line of sight. For every mode there exists, at least, one inclination angle which causes the so-called complete cancellation. According to the analysis performed by \cite{Chadid01}, a (\textit{l},\textit{m})~=~(3,1) geometry combined with an inclination angle of about 45\degr should not suffer from these effects. The (\textit{l},\textit{m})~=~(2,1) mode with \textit{i} of about 50\degr should not suffer from complete cancellation either.

It is surprising that the oscillation mode with the highest amplitude in the photometric data is identified as a (\textit{l},\textit{m})~=~(3,1) mode. Partial cancellation as a result of geometrical effects is expected for modes with \textit{l} $\geq$ 3 in photometry \citep{cancellation1}. Considering these geometrical effects, a (\textit{l},\textit{m})~=~(2,1) mode seems more suitable. Additionally, the amplitude distribution is asymmetric and this is a sign that the influence of the other frequencies could not be properly averaged out. As it is not possible to fit an asymmetric amplitude distribution with one mode, we  conclude that the azimuthal number \textit{m}~=~+1 for F1 is constrained but we cannot be sure about the degree \textit{l}. This is in line with the well-known property that the mode identification method in FAMIAS has better capability of identifying \textit{m} than \textit{l} \citep{FAMIAS}.

The results for the oscillation modes associated with the other p-mode frequencies found in spectroscopy varied from line to line, covering a wide range, so unambiguous identification was not possible.

\subsection{The low frequency range}\label{ssec:lowfreq}

\begin{figure}[tb]
\centering
\includegraphics[width=0.49\textwidth,clip]{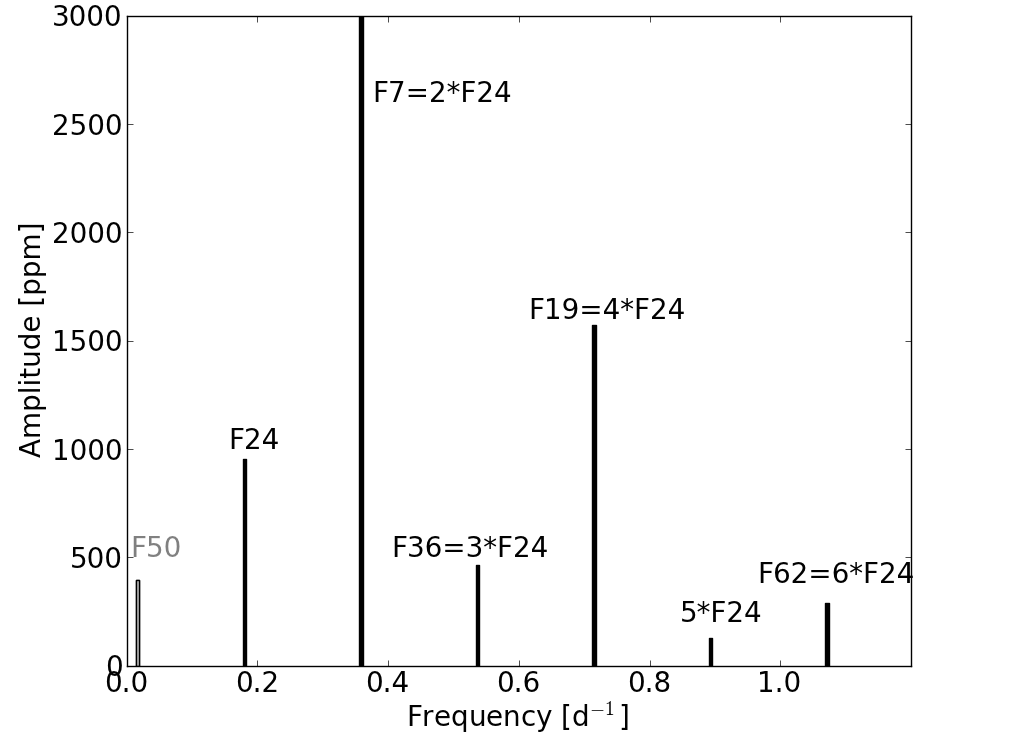}
\caption{Low frequency range of the power spectrum of HD~41641 including the fifth multiple of F24, which we  found did not fulfill the significance criteria.}
\label{lowfreqs} 
\end{figure}

Figure \ref{lowfreqs} shows a zoom into the low frequency peaks obtained from the CoRoT photometric data. Inside the error bars computed for Table \ref{freqs}, the low frequency peaks form a series of harmonics evenly spaced by the value of F24. Only the lowest frequency of the set (F50, plotted in gray) is not part of the series. This frequency was considered to be too small to be pulsation related. It does not appear in the list of known frequencies caused by CoRoT, so it does not seem to be caused by instrumental effects, and it does not match the inverse of the length of the data set (1/T~$\sim$~0.0106~d$^{-1}$). It is related to another effect, which is discussed in Sect.~\ref{ssec:ampl}. The fifth multiple of F24 was missing in the initial frequency list because of a low S/N but was found to have a value of 0.890~$\pm$~0.002 with S/N~$\sim$~3.0. The peak that has the highest amplitude is twice the lowest frequency peak of the series (F7=2*F24).

The phase dispersion minimization (PDM) method, first derived by \cite{pdm}, was also applied to investigate the low frequency range. As was the case when the Fourier technique was applied, the second multiple of the series of harmonics (F7 = 0.3560 $\pm$ 0.0001 d$^{-1}$) was observed to dominate over the first (F24 = 0.17756 $\pm$ 0.00003 d$^{-1}$). The effect of the second harmonic   dominating one of a series found at low frequencies has been observed before in $\delta$~Sct stars (e.g., \citealt{2peak1} or \citealt{2peak2}).

Low frequencies detected in $\delta$~Sct stars outside the p-mode range could have different explanations. They could simply be caused by combinations between high-frequency peaks. However, we did not find combination frequencies that could explain all these peaks within the error bars. In addition, it is unlikely that low-frequency combination peaks appear as a complete series of harmonics. These frequencies could also be g-modes, i.e., typical $\gamma$~Dor frequencies, which would make the target a hybrid pulsator instead of a pure $\delta$~Sct star. Many A- and F-type stars observed by Kepler have been proven to be hybrids (e.g., \citealt{hybrids2}; \citealt{hybrids}; \citealt{VanReeth15}). In our case, the clear harmonic structure, characterized by odd ($2n$ with $n$=1,2,3) harmonics with amplitudes larger than the respective even ($2n-1$) harmonics, more likely supports another possible origin: a double-wave light curve caused by binarity and/or stellar spots.

\begin{figure}[tb]
\centering
\includegraphics[width=0.49\textwidth,clip]{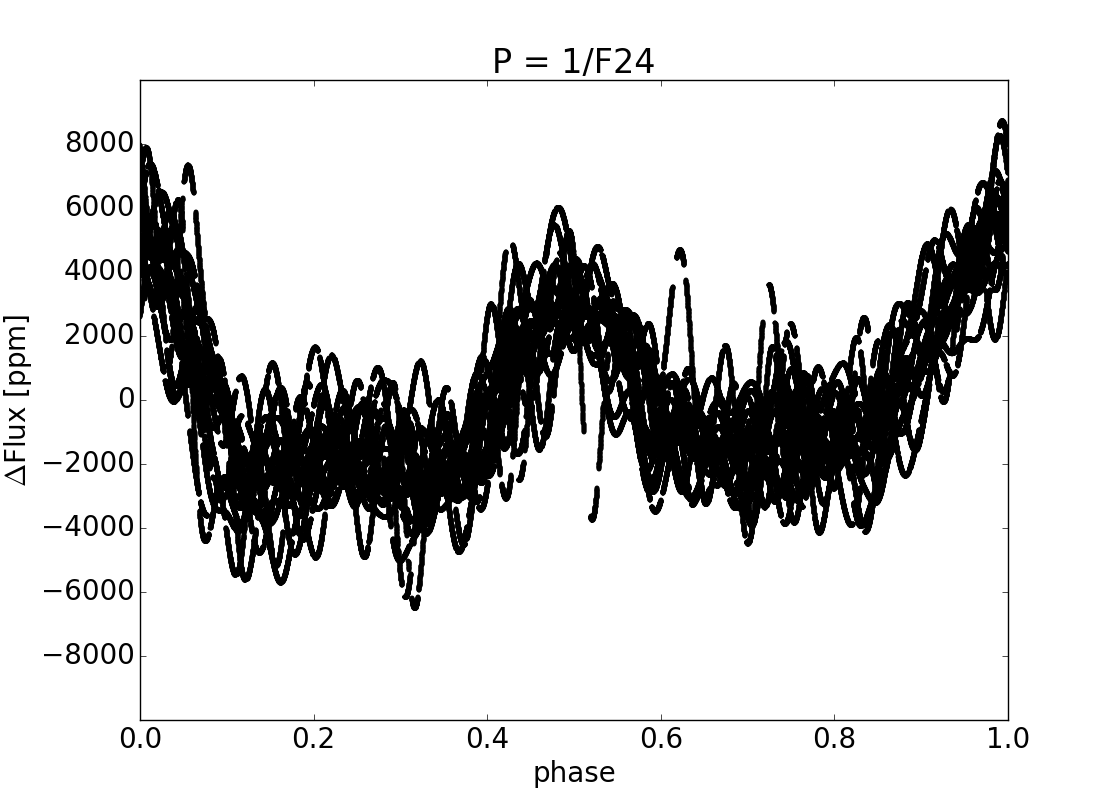}
\caption{\label{foldedLC} Photometric data folded in phase with the period P = 1/F24 after having prewhitened the light variations caused by pulsations.}
\end{figure}

\begin{figure}[tb]
\centering
\includegraphics[width=0.49\textwidth,clip]{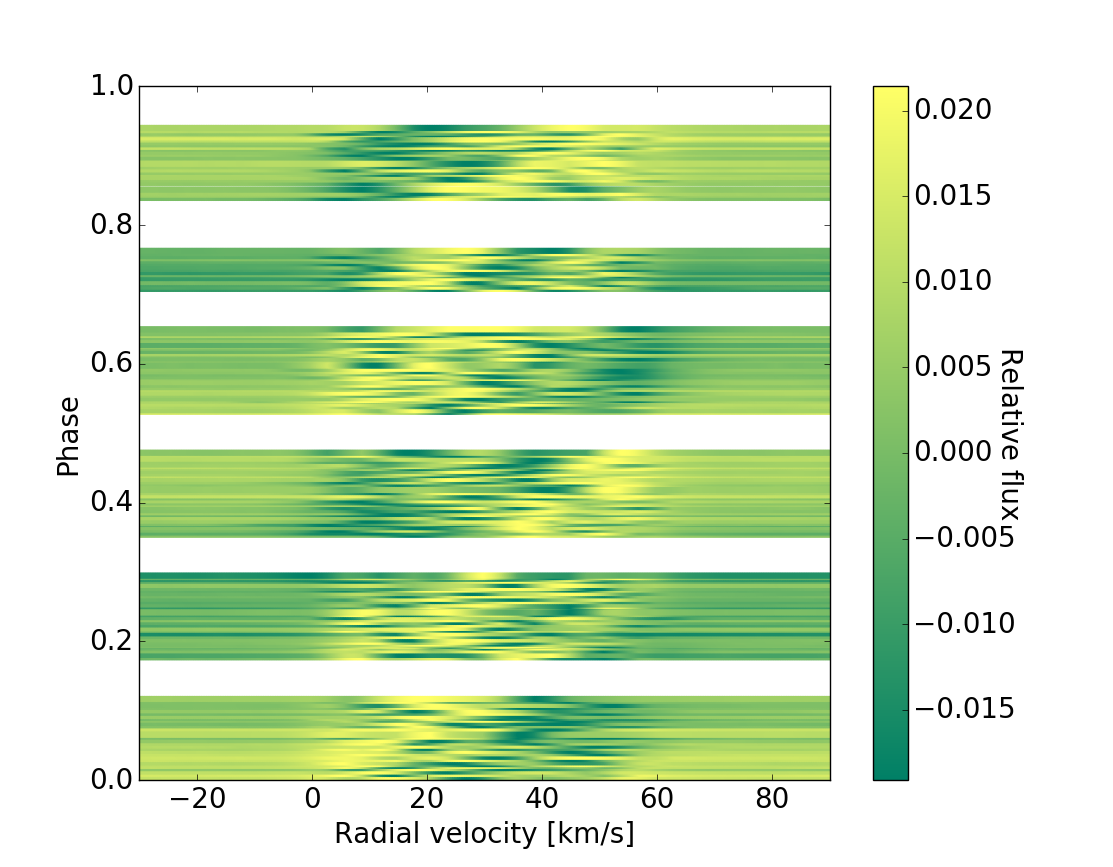}
\caption{\label{LSD} Intensity diagram of the residual flux after subtracting the mean profile of each LSD profile, folded in phase with the period P = 1/F24.}
\end{figure}

\begin{figure*}[tb]
\centering
\includegraphics[width=0.65\textwidth,clip]{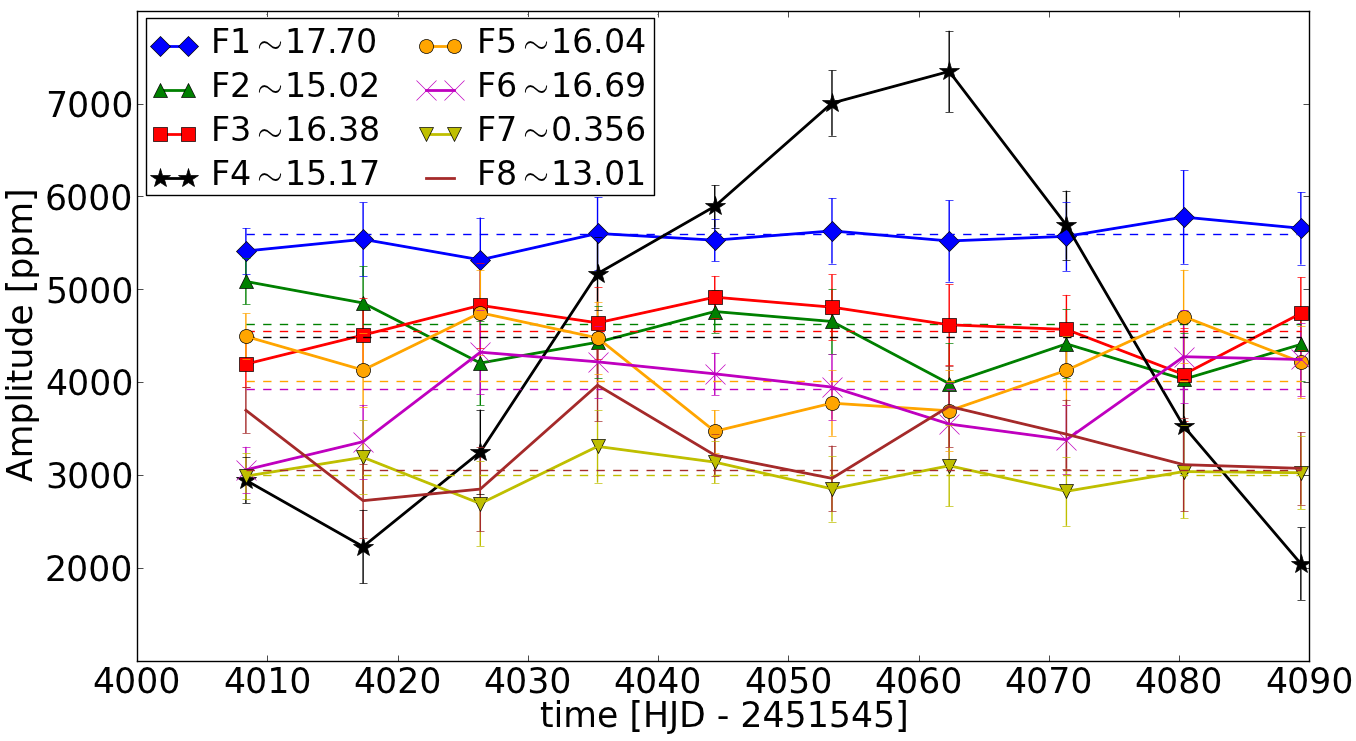}
\caption{\label{AmplVar} Study of the amplitude variability of the eight dominant frequency peaks. The variability of F4 is not instrinsic, but due to the
unresolved frequency F11.}
\end{figure*}

In the case that the star had a noneclipsing orbital companion, periodic orbital modulation effects would create variability in the light curve. \cite{ThisStar} suggested that HD~41641 could be a double-lined spectroscopic binary. On the other hand, dark (bright) spots on the surface would cause periodic reductions (increases) in the detected flux related to the rotation frequency. In both cases, nonlinear features that would appear in the light curve could not be described  with a sinusoidal function well, and the presence of the harmonics would be justified. It is not trivial to distinguish between these two possible scenarios.

A low-pass band filter was applied to the light curve to prewhiten the high frequencies. The contribution of  low frequencies was then phase folded with F24 and  is shown in Fig. \ref{foldedLC}. The observed features cannot be described with a single sine curve, which explains the presence of the series of six detected harmonics. This signature is not typical for a binary light curve but rather points to a spotted structure as the cause of F24.

We also looked for evidence of binarity on the HARPS spectra by applying a least-squares deconvolution analysis (LSD; see \citealt{Donati97} for the original formulation of the method). In an LSD analysis, an average line profile is created that represents each spectrum assuming that all its lines are similar in shape and can be represented by the same profile scaled in depth by a certain factor. It is also assumed that the intensities of overlapping spectral lines add up linearly. This technique increases the S/N enormously and makes it easier to study the line profile variations. More details about the extraction of the LSD profiles can be found in \cite{LSDref}.

The HJD of each spectrum was converted into phase using F24 = 0.17756 d$^{-1}$. In Fig.~\ref{LSD}, residual fluxes with respect to the average line profile are plotted, covering the phase in an intensity diagram. We find that there is no radial velocity shift of the profiles following the phase, as would be expected if HD~41641 were a binary star. The same analysis was repeated using F7 and yielding the same results.

As a final test for binarity, spectral disentangling in Fourier space was performed using the FDBinary code \citep{Ilijic04}. This method enables the separation of the individual spectra of the components of binary and multiple systems, along with the simultaneous determination of their orbital elements. In the case of HD~41641, no second spectrum was detected, allowing us to conclude that HD~41641 does not have an orbital companion. We therefore argue that the presence of spots on the stellar surface creates line profile variations as well as  flux variations that were observed in the light curve. These changes are periodic with the rotation frequency.

It is difficult to unambiguously identify which peak corresponds to the rotation frequency. The F7 peak has higher amplitude in the Fourier spectrum and dominates the PDM. However, using the derived value of \textit{v}sin\textit{i} and the rotation frequency candidates, we can derive the inclination angle, and F24 gives $i \sim$ 45$^o$, which agrees with the values obtained with FAMIAS for the two most probable mode geometries. We also checked that neither F7 nor F24 would imply a rotation velocity close to the critical breakup velocity. Even considering limiting cases, F7 would imply a rotation velocity of about 26$\%$ of the critical velocity, and F24 would still be less than 50$\%$.

Our analysis of the individual chemical abundances suggests that the spots could be chemical, and this could point to the presence of a magnetic field. Spectropolarimetric data are being assembled to test this hypothesis. In order to present a more detailed analysis of the spot distribution, we unsuccessfully attempted Doppler imaging (e.g., \citealt{DoppIm}). From Fig.~\ref{LSD}, it can also be observed that the short period oscillations (i.e., the p-mode frequencies) have high amplitudes over the rotational phase. This influence of the pulsations complicated a deeper study of the surface structures.

\subsection{Amplitude variability}\label{ssec:ampl}

From the study of the complete light curve, 90 frequencies were detected. It is possible that these detected frequencies change amplitude over time and the temporal amplitude variability was studied  to test this effect. Amplitude variability has previously been observed in space data of $\delta$~Sct stars (e.g., \citealt{Bowman15} and \citealt{modulationnew}).

The light curve was divided in ten subsets of ten days in length with an overlap of one day with previous and subesquent subsets. Naturally, fewer frequencies were detected in comparison to the analysis of the complete data set. Only the eight dominant frequencies of the complete set appeared as significant frequencies in the ten sections. Figure~\ref{AmplVar} shows the measured amplitudes of these frequencies against the central time of each subset. Different colors and symbols are used for each frequency, and the dashed horizontal line of the same color corresponds to the amplitude measured for the complete data set. Most of the frequencies vary slightly inside the error bars, but F4 presents very strong variations over time. This amplitude variability is due to the close frequency F11, which is unresolved from F4 over a time baseline of only ten days. Indeed, we observed the expected  beat period of 76 days  between the two frequencies (Fig.~\ref{AmplVar}). We notice that a component close to the F4-F11=0.01319~d$^{-1}$ difference has been detected in the solution of the whole data set (F50=0.01396~d$^{-1}$, Table 4). The discrepancy between the F50 and F4-F11 values suggests a mode interaction or the excitation of another mode close to the resonance between F4 and F11.

\section{Summary}
We carried out an observational asteroseismic study of the $\delta$~Sct star HD~41641. This was based on two different data sets that had been simultaneously obtained: a 94-day photometric data set obtained with the CoRoT space telescope between December, 2010 and March, 2011, and 222 spectra obtained with the HARPS high-resolution spectrograph between December 23, 2010 and January 12, 2011.

The spectroscopic data were used to derive the fundamental stellar parameters of HD~41641. The star was determined to have an effective temperature of 7200 K, a surface gravity of 3.5 dex, a metallicity of -0.2 dex compared to solar, and to rotate with a projected velocity of 30 $\mathrm{km\,s}^{-1}$. Also the microturbulent velocity was determined to be low, $\xi$ = 1.1 $\mathrm{km\,s}^{-1}$. These values also allowed us to estimate the position of HD~41641 in a (\ensuremath{\log g}-log$T_\mathrm{eff}$) diagram and, hence, to put some constraints on its mass and radius. Additionally, the individual chemical abundances were derived, and HD~41641 was found to present signatures that suggest it is an Ap star, although it shows pure $\delta$~Sct pulsations. Ap stars are magnetic chemically peculiar stars, and our analysis, thus, allows us to suggest that HD~41641 might present a magnetic field as well. We plan to perform and analyze spectropolarimetric measurements of HD~41641 as soon as possible to test this hypothesis and evaluate the possible cause of the  rotationally modulated signal we detected. At present, the classical $\delta$~Sct pulsator HD~188774 is the only target proven to have a magnetic field \citep{magdelta}, while the Ap classification and the presence of the magnetic field in the $\delta$~Sct star HD~21190 (\citealt{Koen01}; \citealt{Kurtz08}) are questioned and have to be revisited with a detailed spectroscopic and spectropolarimetric study \citep{Bagnulo12}.

The frequency analysis performed with the CoRoT light curve revealed a total of 90 significant frequencies. All of these frequencies were detected in a range between 0 d$^{-1}$ and 40 d$^{-1}$ and distributed in three groups of frequencies. The group of peaks detected at the lowest frequency domain was found between 0 d$^{-1}$ and 1.2 d$^{-1}$. We concluded that they are not oscillation frequencies but associated with rotational modulation owing to spots on the stellar surface. This was chosen as the most likely scenario as we could not find any evidence of binarity. We did not detect Doppler shift in our study of the line profile variation, and the spectral disentangling technique did not show the presence of two objects either. We also ruled out the possibility that HD~41641 is a hybrid pulsator. The low frequency peaks clearly form a series of harmonics and this structure more likely points to another scenario. Additionally, the derived abundance pattern could be related to the formation of chemical spots and, hence, to rotational modulation of the data. A similar harmonic structure was detected in the case of HD~188774 \citep{2peak1}.

The other two groups of peaks found in the photometric analysis are within the expected p-mode frequency range for a pure $\delta$~Sct star. All the peaks identified in the group with higher frequencies were concluded to be combination frequencies and not independent oscillation frequencies of the star. Hence, the independent p-mode frequencies of HD 41641 lie between 10 d$^{-1}$ and 20 d$^{-1}$.

Finally, a frequency analysis was conducted on the spectroscopic data set. Periodicities were searched using nine metal lines and their integrated moments. Most of the found frequencies had also been detected during the photometric analysis. However, we found much fewer frequencies  and a clear dominance of the low frequencies complicated the mode identification. We could only constrain the azimuthal order \textit{m}~=~+1 of a nonradial prograde mode associated with F1. The inclination angle associated with the most probable pulsation mode geometries is consistent with F24~=~0.17756~$\pm$~0.00003~d$^{-1}$ being the rotation frequency.

\begin{acknowledgements}
A.E. is grateful to the whole IvS team for the enriching discussions and valuable feedback. Additionally, Prof. Dr. Hans Van Winckel, Dr. Lionel Siess, and Prof. Dr. Alain Jorissen are acknowledged. This research was funded by the Belgian Science Policy Office under contract BR/143/A2/STARLAB. K.Z. acknowledges support by the Austrian Fonds zur F\"{o}rderung der wissenschaftlichen Forschung (FWF, project V431-NBL). T.R. acknowledges partial financial support from the Presidium of RAS Program P-41. E.P. and M.R. acknowledge financial support from the FP7 project SPACEINN: Exploitation of Space Data for Innovative Helio-and Asteroseismology.
\end{acknowledgements}

\bibliography{literature}

\end{document}